\documentclass[epj]{svjour}
\RequirePackage{float}
\usepackage{epstopdf}
\usepackage{hyperref}
\usepackage{capt-of}
\usepackage{graphicx}  
\usepackage{dcolumn}   
\usepackage{bm}%

\begin{document}
\title{Temporally Varying Universal Gravitational ``Constant" and Speed of Light in Energy Momentum Squared Gravity}
\author{S. Bhattacharjee\inst{1} \and P.K. Sahoo\inst{2}
}                     
%
%
\institute{Department of Astronomy, Osmania University, Hyderabad-500007,
India,  Email: snehasish.bhattacharjee.666@gmail.com \and Department of Mathematics, Birla Institute of
Technology and Science-Pilani, Hyderabad Campus, Hyderabad-500078,
India,  Email:  pksahoo@hyderabad.bits-pilani.ac.in}
\date{Received: 27th Aug. 2019 / Accepted: date}
%
\abstract{
Energy Momentum Squared Gravity (EMSG) [M. Roshan \& F. Shojai, Phys. Rev. D, \textbf{94}, 044002, (2016)] is a cosmological model where the scale factor is non vanishing at all times and hence does not favor big bang cosmology. However, the profile of density in the radiation dominated universe shows that EMSG supports inflationary cosmology. Inflationary cosmological models are successful in providing convincing answers to major cosmological issues like horizon problem, flatness problem and small value of cosmological constant but hitherto no model of inflation has been observationally confirmed. Owing to this, Varying Speed of Light (VSL) were introduced which are a class of cosmological models which disfavor inflation and propose an alternative route to solve these cosmological issues by just allowing the speed of light (and Newtonian Gravitational constant) to vary. VSL theories were motivated to address the shortcomings of inflation but do not address the shortcomings related to the initial big bang singularity. In this spirit, we present here a novel cosmological model which is free from both the ``initial big bang singularity" and ``inflation" by incorporating a mutually varying speed of light $c(t)$ and Newtonian gravitational constant $G(t)$ in the framework of EMSG. We report that in EMSG, for a dust universe ($\omega=0$), cosmological models for a time varying $c(t)$ and $G(t)$ and constant $c$ and $G$ are indistinguishable, whereas for a radiation dominated universe ($\omega = 1/3$), a mutually varying $c(t)$ and $G(t)$ provides an exiting alternative to inflationary cosmology which is also free from initial big bang singularity. We further report that for an ansatz of scale factor representing a bouncing cosmological model, the VSL theory can be applied to a quadratic $T$ gravity model to get rid of ``inflation" and ``big bang singularity" and concurrently solve the above mentioned cosmological enigmas.
} 
\titlerunning{Temporally Varying Universal Gravitational... in EMSG} \authorrunning{S. Bhattacharjee, P.K. Sahoo}
\maketitle
\section{Introduction}
Varying speed of light (VSL) theories refer to the lineage of hypothesis that speed of light is spatially, temporally, wavelength and frequency dependent. The first VSL theory was  proposed by Einstein \cite{e1,e2} where he concluded that speed of light is a constant only in the vicinity of constant gravitational potential or zero gravitational influence. Einstein stated that speed of light is a function of frequency and position and in accordance to Huygen's principle \cite{huy}, light rays traveling normal to any gravitational field must confront curvature \cite{e2}. Subsequently in 1957, R. Dicke proposed speed of light being a function of wavelength in addition to position and frequency and inferred that VSL theories can provide alternate explanation to cosmological redshift \cite{dicke}. In Quantum Field Theory (QFT), Heisenberg uncertainty principle \cite{hei,h2,h3} do permit virtual photons traveling faster than real photons for short interim which does not violate causality as they do not transfer information\cite{feyn}. In recent times, A. Albrecht \& J. Magueijo proposed light traveling many orders of magnitude faster in the early universe, when a sudden phase transition of unknown origin dropped its speed to the currently observed value \cite{mag}. VSL theories have been proclaimed as an alternate to inflationary cosmology and provide viable solutions to flatness problem, horizon problem  and small value of the cosmological constant \cite{19,20,21,22,moffat,barrow,mag,mag2000,27,28,29}. \\
In addition to the speed of light, other physical constants are also posited to be functions of time. In 1937, P. Dirac argued temporal variations of universal gravitational constant of about 5 parts in $10^{11}$ per year can explain the relative small strength of the gravitational force compared to all other fundamental forces \cite{dirac}. Webb et al. \cite{a1} affirmed redshift dependency of electromagnetic fine structure constant ($\alpha$) by studying distant quasars while Bekenstein \cite{b,b2} proposed temporal variations in electron charge and reported time varying electron charge could induce variations in $\alpha$.\\
Moreas \cite{moreas} presented cosmological solutions for temporally varying speed of light in the framework of $f(R,T)$ gravity. $f(R,T)$ gravity is a modified gravity theory introduced in the literature by \cite{harko} due to the failure of Einstein's GR \cite{gr} in explaining the present cosmic acceleration without employing mysterious dark energy which hitherto have no observational evidence. Other notable modified gravity theories are $f(R)$ gravity \cite{fr}, $f(\mathcal{T})$ theory \cite{ft} and $f(G)$ gravity \cite{fg}.\\
EMSG is a modified gravity theory recently proposed by \cite{roshan} which is a new covariant generalization of GR allowing the presence of a term proportional to $T^{2} = T_{\mu \nu} T^{\mu \nu}$ (where $T_{\mu \nu}$ is stress-energy-momentum tensor) in the action in addition to the Ricci scalar $R$. Cosmological solutions in EMSG for isotropic and anisotropic space-times have been reported in \cite{cs}. The paper is organized as follows: In Section II we present an overview of EMSG and derive the field equations for a FLRW metric. In Section III we present cosmological solutions for  varying $c(t)$ and $G(t)$ in matter and radiation dominated universes. In Section IV we present an alternate route to investigate the validity of the proposed cosmological model. Finally in Section V we present our results and conclude the work.

\section{Overview of Energy momentum squared gravity}
The action in EMSG is given by 
\begin{equation}\label{1}
S=\frac{c^{4}}{16\pi G} \int \sqrt{-g}\left[R - \zeta T^{2} - 2 \Lambda + \mathcal{L}_{m}\right] dx^{4}
\end{equation}
where $\mathcal{L}_{m}$ is the matter Lagrangian, $\Lambda$ is the  cosmological constant and $\zeta$ is the coupling parameter whose value can be inferred from observations. Positive $\zeta$ values generate acceptable cosmological behaviour \cite{roshan}.
We assume the universe comprising predominantly of a perfect fluid and hence we take $\mathcal{L}_{m}=p$.\\
Varying the action \ref{1} with respect to metric yields the field equations as 
\begin{equation} \label{2}
\mathcal{G}_{\mu \nu} = \frac{8 \pi G}{c^{4}} T^{eff}_{\mu \nu} - \Lambda g_{\mu \nu}
\end{equation}
where $\mathcal{G}_{\mu \nu}$ represents the Einstein tensor and $T^{eff}_{\mu \nu}$ is given by 
\begin{equation} \label{3}
T^{eff}_{\mu \nu}  - T_{\mu \nu}= \frac{\zeta c^{4}}{4 \pi G}\left( \Phi _{\mu \nu} - \frac{1}{4}g_{\mu \nu} T^{2} + T^{\sigma}_{\mu}T_{\mu \nu}\right) 
\end{equation}
where 
\begin{equation}
\Phi _{\mu \nu} = T^{\alpha \beta} \frac{\delta T_{\alpha \beta}}{\delta g ^{\mu \nu}}
\end{equation} 
We take $T_{\mu \nu} = \left( p + \rho \right)u_{\mu}u_{\nu} + p g_{\mu \nu} $, where $p$ denote pressure and $\rho$ denote density.\\
For a flat FLRW metric with ($-$,$+$,$+$,$+$) metric signature we obtain the following Friedman equations 
\begin{equation}\label{4}
H^{2} = \frac{8 \pi G}{3 c^{4}} \rho - \zeta\left[\frac{1}{2} p^{2} + \frac{1}{6} \rho^{2}  + \frac{4}{3} \rho p\right] +\frac{\Lambda}{3}
\end{equation}
\begin{equation}\label{5}
\frac{\ddot{a}}{a}= -\frac{4 \pi G}{3 c^{4}}\left( \rho + 3p\right) + \zeta\left[ p^{2} + \frac{1}{3} \rho^{2}  + \frac{2}{3} \rho p \right] +  \frac{\Lambda}{3}
\end{equation}
where $H$ represents Hubble parameter, $a$ denote scale factor and overhead primes symbolize time derivatives. 
\section{cosmological solutions in EMSG for $c\rightarrow c(t) \&   G \rightarrow G(t)$}
We will now present cosmological solutions for \ref{4} and \ref{5} for matter and radiation dominated universes.\\
To incorporate VSL theory into EMSG, we will have to make a prior assumption of how the speed of light and Newtonian Gravitational constant vary with time. Since it is unclear whether these physical quantities vary with time and if they do what is their dependency on time, we will be following the prescription of \cite{barrow} by assuming power-law evolution of speed of light $c(t)$ and gravitational constant $G(t)$  with respect to scale factor as 
\begin{equation}\label{6}
c(t)= c_{0} a(t)^{r}
\end{equation}
\begin{equation}\label{7}
G(t) = G_{0} a(t) ^{s}
\end{equation}
where $c_{0}$ and $G_{0}$ represents current value of these quantities. We plan to work with natural units, thence we set $c_{0} = 8 \pi G_{0} = 1$.

\subsection{Matter dominated Universe ($p=0$)}
Substituting $p=0$ in \ref{4} \& \ref{5} expression of scale factor $a(t)$ reads \cite{cs} 
\begin{equation}\label{8}
a(t)= 4\left[( \epsilon^{2} + \varepsilon + 1) \cosh\left(\sqrt{\frac{3\Lambda t}{2}} \right) + ( \epsilon^{2} + \varepsilon - 1)\sinh\left(\sqrt{\frac{3\Lambda t}{2}} \right) - 2\epsilon \right]^{1/3} \Lambda^{-1/3}
\end{equation}
where $\epsilon$ \& $\varepsilon$ are constants.\\
Using \ref{8} one easily obtain an expression for deceleration parameter $q= -\frac{a\ddot{a}}{\dot{a}^{2}}$. 
\begin{figure}[H]
\begin{center}
\includegraphics[width=7.75cm]{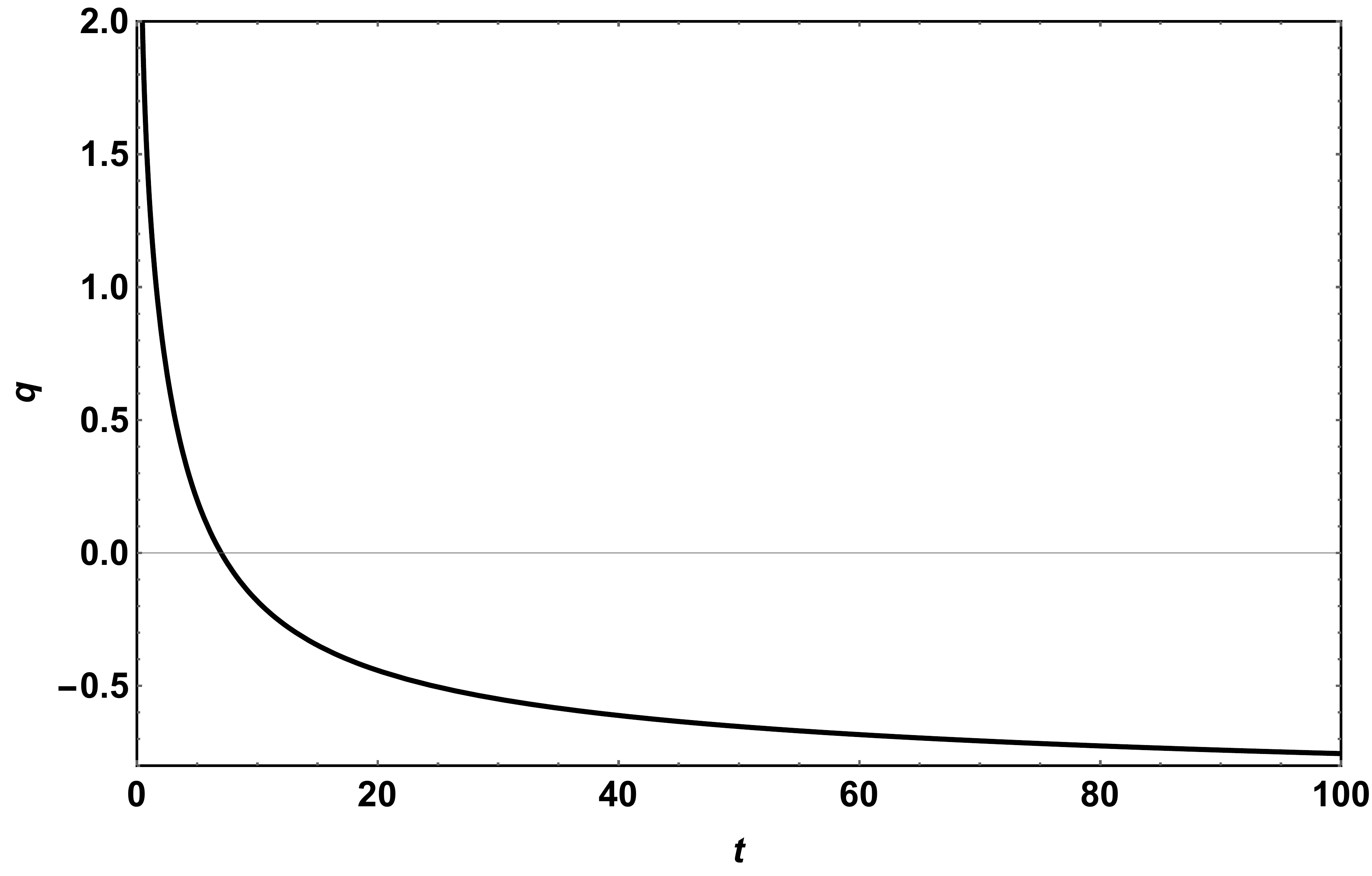}
\caption{Time evolution of deceleration parameter ($q$) in matter dominated universe with $\epsilon = \varepsilon = \Lambda =1$}\label{f1}
\end{center}
\end{figure}
Time evolution of $q$ is shown in Figure \ref{f1} where we observe traces of decelerated expansion in the early history of matter dominated universe. As a matter of fact, one expects a decelerated expansion ($q>0$) in the context of $\Lambda$CDM models derived from GR. Nonetheless, $q<0$ at later times which is in agreement with current observations. This is remarkable in the sense that we have assumed $p=0$ and still we obtain an accelerated expansion in the framework of EMSG. 
Expression of density $\rho$ is given as \cite{cs}
\begin{equation}\label{15}
\rho = C a^{-3}
\end{equation}
where $C$ is a constant. From \ref{15}, it is clear that the expression of density is insensitive to time varying $c(t)$ and $G(t)$. Hence time varying $c(t)$ and $G(t)$ does not add any new information for a dust universe in EMSG. 
\subsection{Radiation dominated Universe ($p=\rho / 3$)} 
\subsubsection{Time varying $c(t)$ \& $G(t)$}
Substituting  $p=\rho / 3$ in \ref{4} \& \ref{5},  expression of scale factor $a(t)$ becomes \cite{roshan} 
\begin{equation}\label{9}
a(t)= A \sqrt{\cosh Bt}
\end{equation}
where $A$ is a constant and $B = \sqrt{4 \Lambda/3}$.\\
The deceleration parameter in this case reads 
\begin{equation}\label{10}
q(t)= 1 -\tanh(Bt)^{2}
\end{equation}
which condenses to $1$ for $t=0$. This is strictly the value of deceleration parameter for a radiation dominated universe in GR. However $q<0$ for all $t>0$.\\
Expression for density ($\rho$) reads 
\begin{equation}\label{11}
\rho (t, r, s, \zeta) = \frac{\left( A \sqrt{\cosh (Bt)}\right)^{s - 4r} \left[ 1 + \left(1 + 8 \zeta   A \sqrt{\cosh (Bt)}   \right)  \right]^{8r-2s} }{4 \zeta}
\end{equation}
From \ref{11} it is clear that expression of density is sensitive to time varying $c$ and $G$. Figure \ref{f2} illustrates time evolution of density ($\rho$) which clearly supports non-inflationary scenario. This is because VSL theories postulate no traces of super-exponential expansion in the early history of the universe, which should be realized by a rapid decrease in energy density of the universe steeper than standard big-bang cosmology. This is clearly depicted in Figure \ref{f2}.\\
In Figure \ref{f2}, we set $r=-2.5$, $s=-11$ \& $\Lambda = A =  \zeta = 1$. We note that the choice of $r$ and $s$ are not chosen at random. We appointed these values in order to resolve some of the most outstanding cosmological issues such as flatness problem, horizon problem and small value of the cosmological constant. As reported in \cite{barrow}, for matter following an equation of state (EoS) of the form $\rho (\eta - 1)  = p$, the horizon problem retires as long as 
\begin{equation}\label{12}
r\leq 0.5 (2 - 3 \eta)
\end{equation}
is satisfied. In a radiation dominated universe, $p = \rho /3$ \& hence $\eta = 4/3$. Consequently, $r$ follows the restriction \begin{equation}
r\leq -1
\end{equation} In \cite{barrow} it was emphasized that time variation of $G$ does not influence the restriction imposed on $r$ in a significant way. Ergo, the demarcation on $r$ is valid for both a constant or a time varying $G$. \\
The constraint on $r$ to expound the flatness problem is exactly same as horizon problem. Here $r$ has to obey the restriction 
\begin{equation}\label{13}
2 r \leq 2 - 3 \eta
\end{equation}          
which for a radiation dominated universe is again $r\leq -1$. Similar to the horizon problem, flatness problem does not require a time dependent $G$. Hence \ref{7} have no impact in solving both the flatness and horizon problems. A time varying $c$ as in \ref{6} is adequate in resolving these cosmological issues.\\
However a time varying $G(t)$ joins the limelight with $c
(t)$ when one tries to resolve the small value of the cosmological constant at the present epoch in VSL theories \cite{barrow}. For a mutually varying $G(t)$ and $c(t)$, the cosmological constant problem gets thrown out of the window as long as 
\begin{equation}\label{14}
r\leq \frac{-3}{2} \eta
\end{equation}  
is satisfied, which for a radiation dominated universe is $r \leq -2$. We report an additional constraint on $s$ in order to obtain very rapid dilution of energy density in radiation dominated universe as
\begin{equation}
s - 4 r < 0
\end{equation} 
Hence our choice of $r= -2.5$ and $s= -11$ satisfies \ref{12}, \ref{13}, \ref{14} and concurrently straightens out the above mentioned cosmological issues revoking all inflationary scenarios.\\
\subsubsection{Constant $c$ \& $G$}
We now investigate the case where the speed of light $c$ and Newtonian gravitational constant $G$ are held constant. For such a cosmological model, the expression of density reads \cite{roshan}
\begin{equation}
\rho (t, \zeta) = \frac{\kappa}{4 \zeta}\left(1+\sqrt{1+\frac{8 \zeta \Lambda}{\kappa^{2}}sech^{2} B t} \right) 
\end{equation}
\begin{figure}[H]
\begin{center}
\includegraphics[width=8cm]{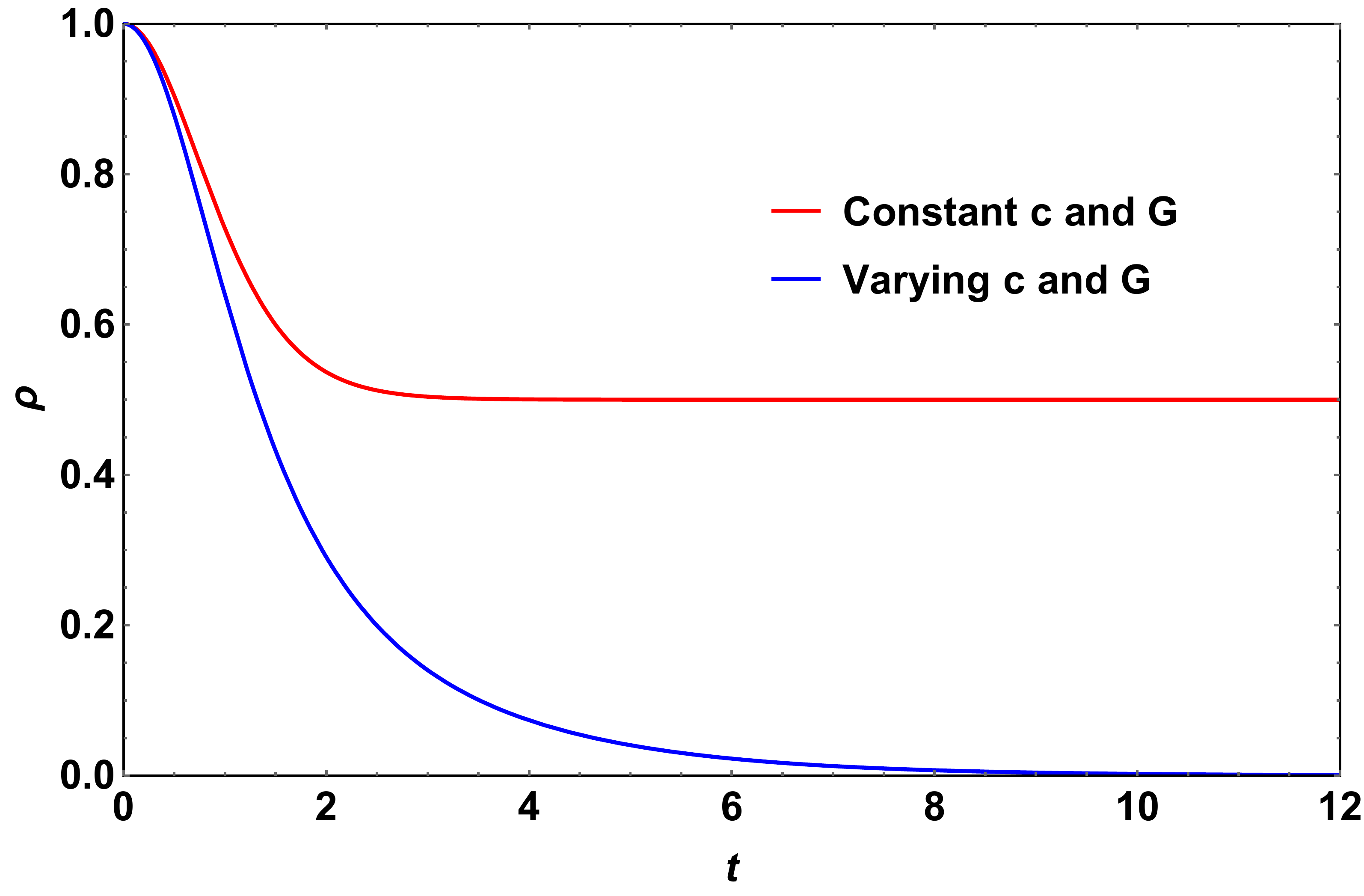}
\caption{Time evolution of density ($\rho$) for constant and varying $c$ and $G$ cases in radiation dominated universe}\label{f2}
\end{center}
\end{figure}
In Figure \ref{f2}, we observe that for constant $c$ and $G$, the density dilutes at a rate almost equal to that of the time varying $c(t)$ and $G(t)$ case at early times due to the presence of sech(t) term. However as time passes, the rapid dilution ceases and density decreases very slowly afterwards. An abrupt decrease in density for such a small period of time can be regarded as a reminiscent of inflationary cosmology. 
\section{validation of the EMSG-VSL cosmological model}
\begin{figure}[H]
\begin{center}
\includegraphics[width=8cm]{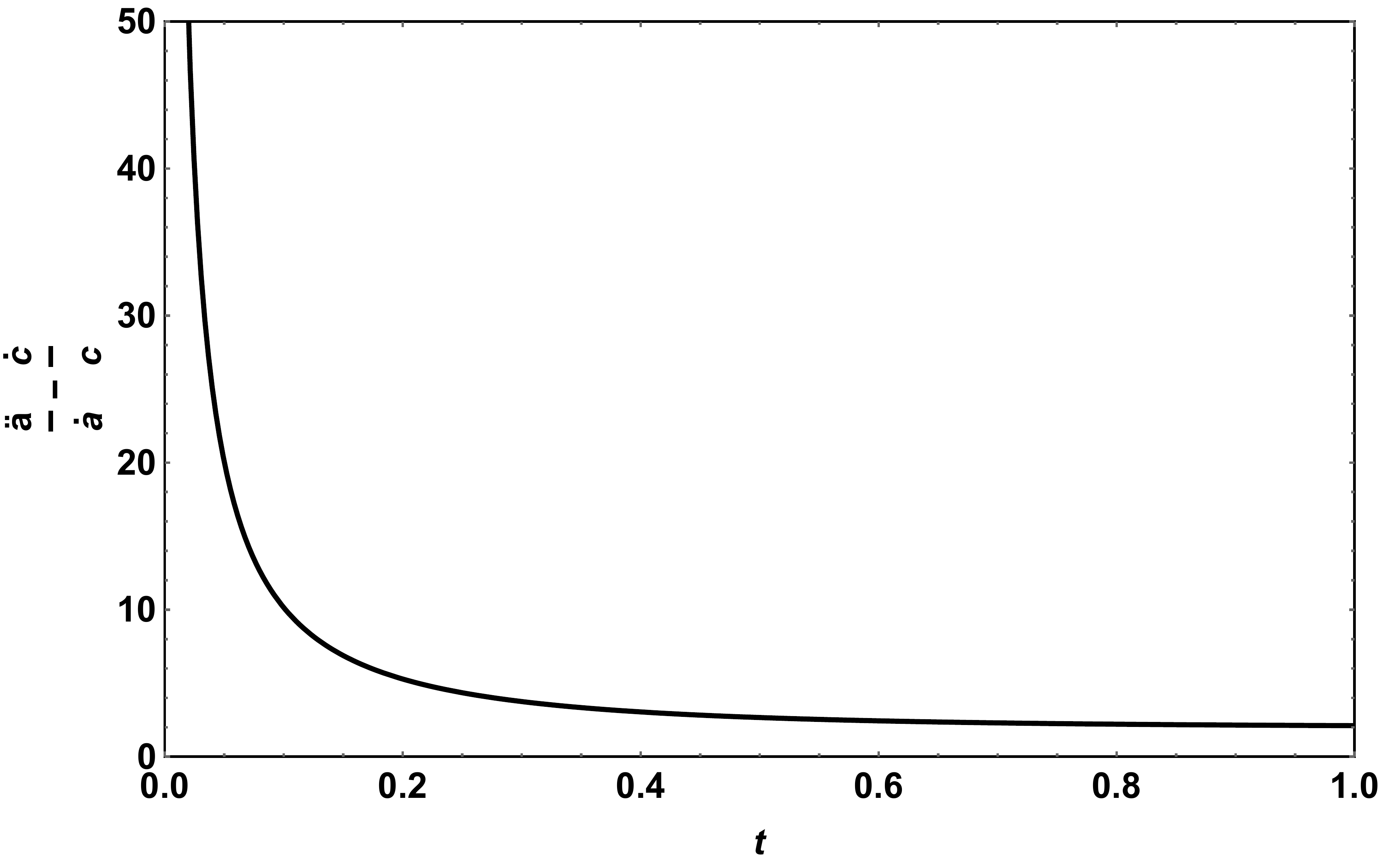}
\caption{Time evolution of $[\ddot{a}/ \dot{a} - \dot{c}/ c]$ in radiation dominated universe with $r =-2.5$ \& $A = \Lambda = 1$ }\label{f3}
\end{center}
\end{figure}

We report an alternative route in verifying the viability of EMSG with a mutually varying $G(t)$ and $c(t)$ in solving the above addressed cosmological problems.  As we go back into the past, the comoving horizon continuously break down into more comoving casually connected regions, which defines the horizon problem \cite{moreas}. As reported in \cite{mag, moffat2}, a larger value of $c$ could very well connect distant remote places of the universe. The comoving horizon reads $D_{c} = c/ \dot{a}$. Hence, an elucidation of horizon problem and therefore the flatness problem dictate $D_{c}$ to decrease in the past allowing vast cosmic stretches to be mutually connected. Hence as addressed in \cite{mag10}, the criteria 
\begin{equation}\label{100}
0 < \frac{\ddot{a}}{\dot{a}} - \frac{\dot{c}}{c}
\end{equation}
should be followed. In Figure \ref{f3}, we confirm \ref{100} and the profile is indeed restricted to positive values at all $t>0$.

\begin{figure}[H]
\begin{center}
\includegraphics[width=7.75cm]{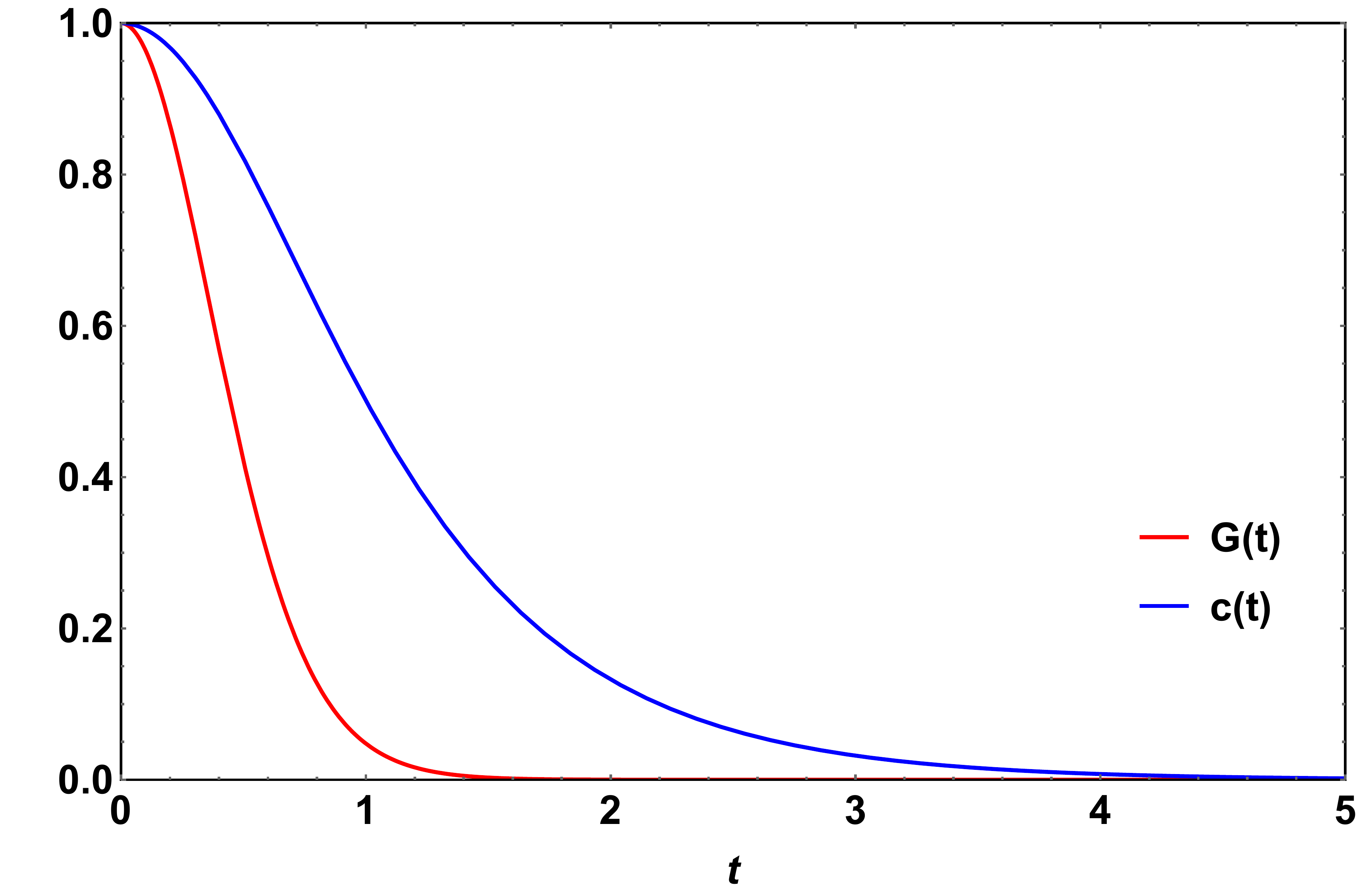}
\caption{Time evolution of $c(t)$ and $G(t)$ in radiation dominated universe with $r =-2.5$, $s=-11$ \& $A = \Lambda = 1$ }\label{f4}
\end{center}
\end{figure}
In Figure \ref{f4}, we present time evolution of $c(t)$ and $G(t)$ for radiation dominated universe. We observe that as reported in \cite{mag}, $G(t)$ and $c(t)$ had very large magnitudes in the early universe compared to their present estimates, while a phase transition of unknown origin dropped their values to the currently observed ones. We highlight that $G(t)$ had to decrease faster than $c(t)$ in order to clear out the cosmological problems addressed here.
\section{Extending the analysis with Quadratic $T$ Gravity}
In this section we present a short analysis of VSL cosmological model in the framework of $f(R,T)$ gravity where $f(R,T) = R + a T + b T^{2}$. Cosmological scenarios for such models have been investigated in \cite{moreas/astro}. For such a model, the action reads \cite{harko}
\begin{equation}
\mathcal{S} = \frac{8 \pi G}{c^{4}} \int \left[ R + a T + b T^{2}\right]\sqrt{-g}d^{4}x 
\end{equation}
The first Friedman equation for such a cosmological model in a FLRW space-time reads \cite{moreas/astro}
\begin{equation}\label{90}
3 \left( \frac{\dot{a}}{a}\right) ^{2} =\frac{8 \pi G}{c^{4}} \rho + \frac{1}{2}\left( a (3 \rho - p) + b (5 \rho + p) (\rho - 3 p)\right) 
\end{equation}
The expression of density $\rho$ for a radiation dominated universe $p= \rho / 3$ then reads
\begin{equation} \label{16}
\rho (t) = \left( \frac{3}{(\frac{8 \pi G}{c^{4}}) + \frac{4 a}{3}}\right) \left( \frac{\dot{a}}{a}\right) ^{2}
\end{equation}
In \cite{moreas/astro}, the authors mentioned that an analytical expression of scale factor does not exist for a radiation dominated universe. Therefore, to proceed in our analysis, we will assume two  ansatze of scale factor. The first ansatz is given as   
\begin{equation}\label{17}
a(t) \sim t^{\gamma}
\end{equation}
where $\gamma > 1$ for an expanding universe. For such a form of scale factor, $a=0$ for $t=0$ and thus is a representative of big bang type cosmological model.
Substituting \ref{6}, \ref{7} \& \ref{17} in \ref{16}, we obtain the expression of density in the radiation dominated universe for a mutually varying $c(t)$ and $G(t)$ as 
\begin{equation}\label{18}
\rho (t, r, s, \gamma) = \left( \frac{3}{8 \pi (t)^{\gamma(s - 4 r)} + \frac{4 a}{3}}\right) \left( \frac{\gamma}{t}\right) ^{2}
\end{equation}
\begin{figure}[H]
\begin{center}
\includegraphics[width=8cm]{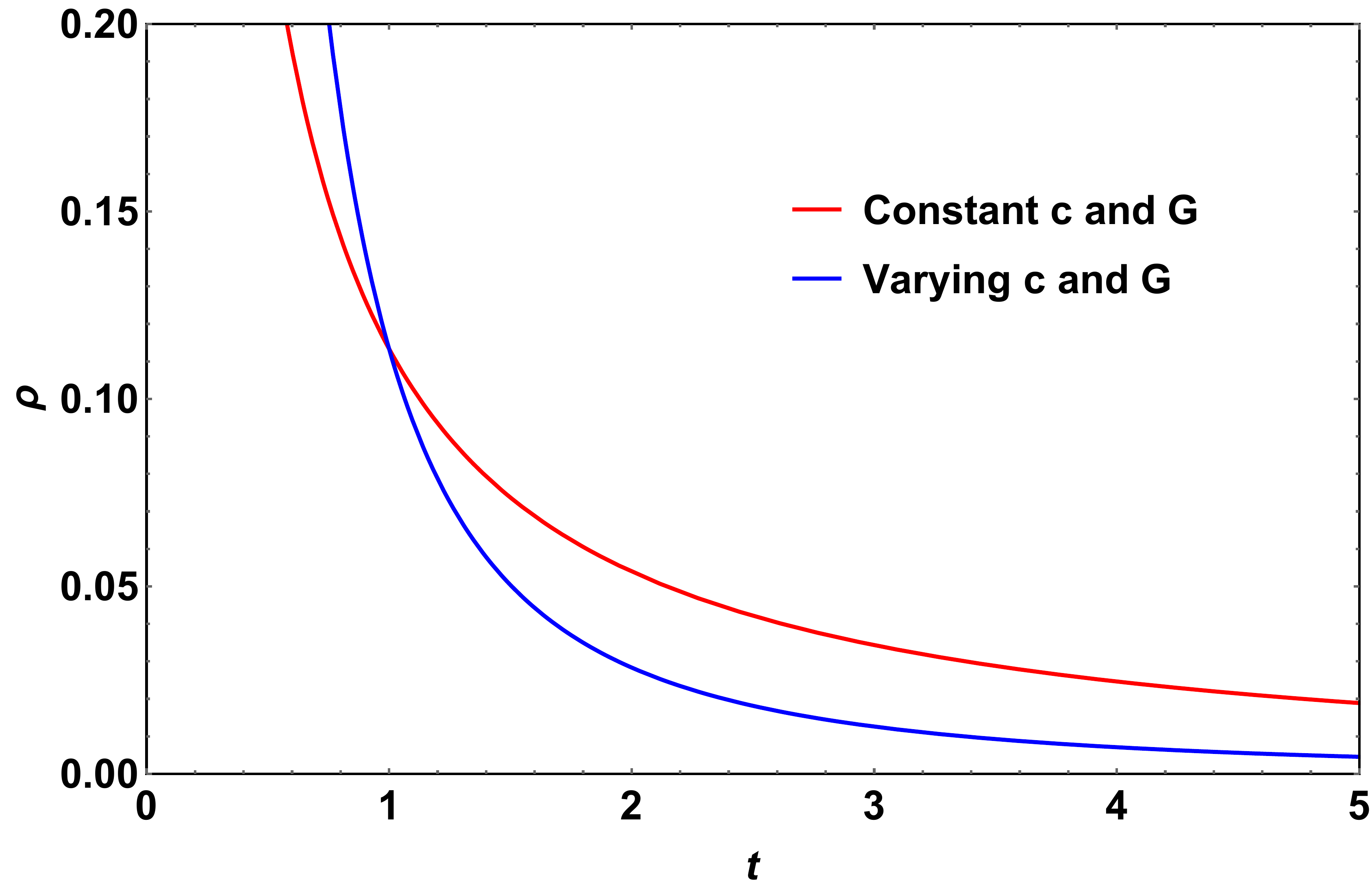}
\caption{Time evolution of density ($\rho$) for constant and varying $c$ and $G$ cases in radiation dominated universe when $a(t) \sim t^{\gamma}$}\label{f5}
\end{center}
\end{figure}
The second ansatz of scale factor is given as 
\begin{equation}\label{19}
a(t) \sim (1 + t^{\gamma})
\end{equation}
This ansatz of scale factor is a representative of bouncing cosmological model since $a > 0, \forall t$. Substituting \ref{6}, \ref{7} \& \ref{19} in \ref{16}, the expression of density $\rho$ in the radiation dominated universe for a mutually varying $c(t)$ and $G(t)$ is given as
\begin{equation}
\rho (t, r, s, \gamma) = \left( \frac{3}{8 \pi (1 + t)^{\gamma(s - 4 r)} + \frac{4 a}{3}}\right) \left( \frac{\gamma t^{\gamma - 1}}{1 + t^{\gamma}}\right) ^{2}
\end{equation}
\begin{figure}[H]
\begin{center}
\includegraphics[width=8cm]{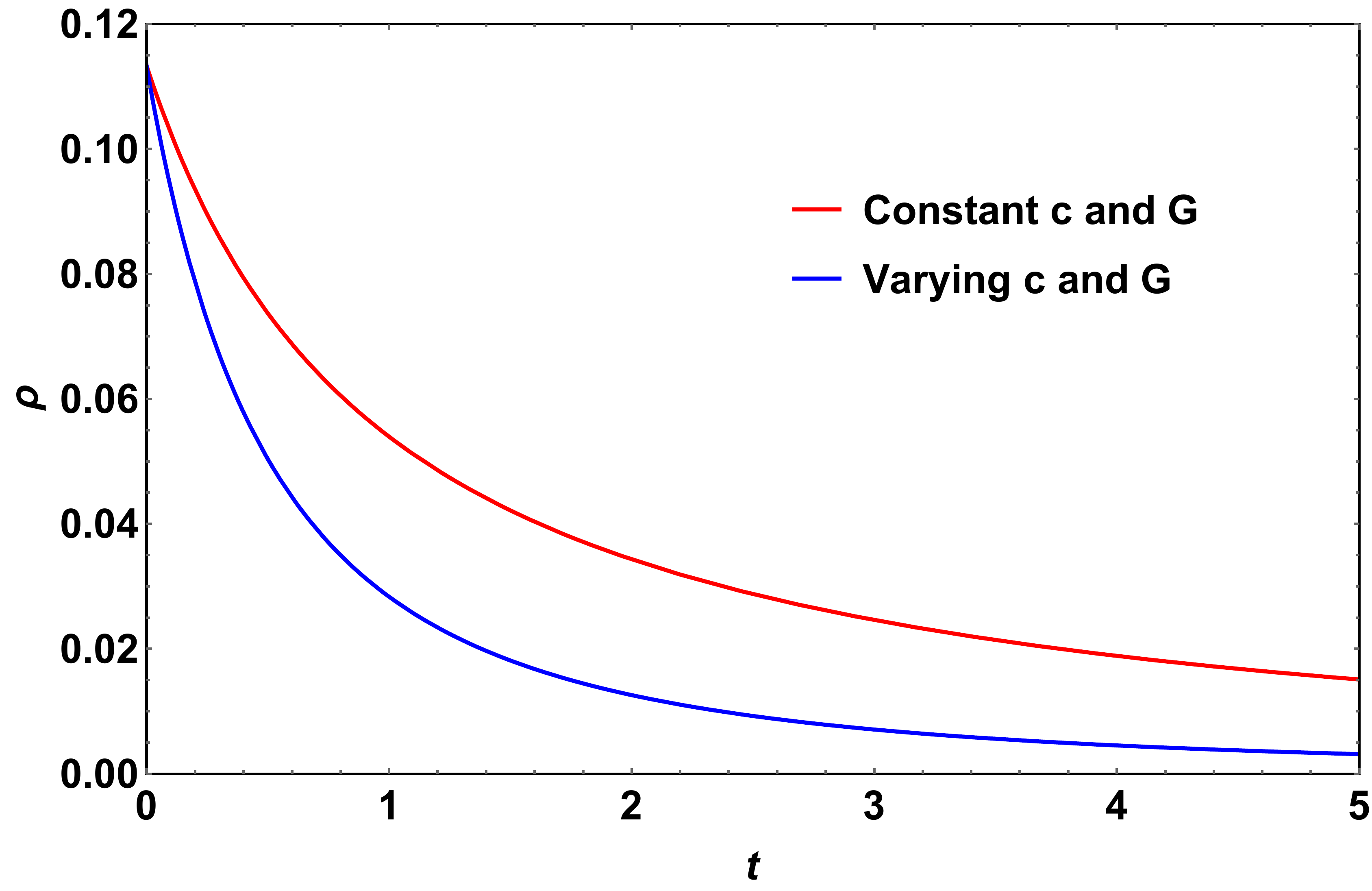}
\caption{Time evolution of density ($\rho$) for constant and varying $c$ and $G$ cases in radiation dominated universe when $a(t) \sim (1 + t^{\gamma})$}\label{f6}
\end{center}
\end{figure}
From Figure \ref{f5} \& \ref{f6}, we clearly observe that density diminishes with time for both constant and varying $c$ and $G$ cases but has a stepper slope for varying $c$ and $G$ for both the ansatze of scale factor. For the second ansatz of scale factor \ref{19}, the VSL theory can be applied to the quadratic $T$ gravity model to get rid of ``inflation" and ``big bang singularity" and concurrently solve the horizon problem, flatness problem and small value of cosmological constant. 

\section{conclusions}
Cosmological models exist where the initial big bang singularity can be avoided by allowing the scale factor to remain finite at all times. Such models are termed ``bouncing cosmological models". These models are constructed in Einstein's gravity and also in many scalar tensor theories of gravity. Energy Momentum Squared Gravity (EMSG) is one such bouncing cosmological model. These models however, do not make an attempt to resolve the shortcomings and problems associated with an inflationary universe. To address these issues, another class of cosmological models were proposed which are termed ``Varying Speed of Light (VSL)" scenarios. VSL models state that major cosmological puzzles like flatness problem, horizon problem and small value of cosmological constant can be resolved without requiring ``inflation" by allowing the speed of light (and Newtonian Gravitational constant) to vary. VSL theories were motivated to address the shortcomings of inflation but do not address the shortcomings related to the initial big bang singularity. In this spirit, we present here a novel cosmological model which is free from both the ``initial big bang singularity" and also ``inflation" by incorporating a mutually varying speed of light $c(t)$ and Newtonian gravitational constant $G(t)$ in the framework of EMSG. We found that for a dust universe ($\omega=0$), cosmological models for a time varying $c(t)$ and $G(t)$ and constant $c$ and $G$ are indistinguishable, whereas for a radiation dominated universe ($\omega = 1/3$), a mutually varying $c(t)$ and $G(t)$ provides an exiting alternative to inflationary cosmology which is also free from initial big bang singularity.\\
We further report that for an ansatz of scale factor representing a bouncing cosmological model, the VSL theory can be applied to a quadratic $T$ gravity model to get rid of ``inflation" and ``big bang singularity" and concurrently solve the horizon problem, flatness problem and small value of cosmological constant.

\textbf{Acknowledgments: }PKS acknowledges CSIR, New Delhi, India for financial support to carry out the Research project [No.03(1454)/19/EMR-II Dt.02/08/2019].

\end{document}